\documentclass[12pt]{article}
\usepackage{amsmath}

\newcommand{\bea}{\begin{eqnarray}}
\newcommand{\eea}{\end{eqnarray}}

\newcommand{\ob}[1]{\left(#1\right)}
\newcommand{\rb}[1]{\left[#1\right]}

\newcommand{\tb}[1]{\left\langle #1\right\rangle }

\title{{\small\hfill SINP/TNP/05-22, IMSc/2005/10/23}\\
\textbf{Gauge dependence of the infrared behaviour of massless
QED$_3$}}

\author{Indrajit Mitra$^{a}$\footnote{indrajit.mitra@saha.ac.in},
Raghunath Ratabole$^{b}$\footnote{raghu@imsc.res.in}~ and
H. S. Sharatchandra$^{b}$\footnote{sharat@imsc.res.in} \\\\
$^a$ Theory Group, Saha Institute of Nuclear Physics, 1/AF Bidhan-Nagar,\\
Kolkata 700064, India\\
$^b$ The Institute of Mathematical Sciences, C.I.T. Campus, Taramani P.O.,\\
Chennai 600113, India} 
\date{}
\begin{document}
\maketitle
\begin{abstract}

Using the Zumino identities it is shown that in a class of non-local
gauges, massless QED$_3$ has an infrared behaviour of a conformal field 
theory with a continuously varying anomalous dimension of the fermion.
In the usual Lorentz gauge,
the fermion propagator falls off exponentially for a large separation,
but this apparent fermion mass is a gauge artifact. 

\end{abstract}
\newpage
%%%%%%%%%%%%%%%%%%%%%%%%%
%%%%%%%%%%%%%%%%%%%%%%%%%

Massless QED in 2+1 dimensions (and in general for space-time 
dimensions $2 < d < 4$) has very interesting features.
It is not just super-renormalizable, it is ultraviolet finite.
The usual perturbation expansion in the fine structure constant 
(which is now a dimensionful parameter) has severe infrared (iR)
divergences which become worse as the number of loops
increases. So QED$_3$ provides an ideal platform for tackling 
iR divergences. This has led to an extensive study of this 
model \cite{jt}.
It is clear that 
the usual perturbation theory can make sense
only by some kind of resummation. 
In a $1/N$ expansion, $N$ being the number of fermion flavours, there is
a resummation of chains of one-loop vacuum polarization 
diagrams on every photon propagator. This changes
the iR
behaviour of the photon propagator from being
inversely quadratic to inversely linear in momentum.
Thus the iR divergence is softened. 
Even after this there are logarithmic iR divergences.
(Throughout this paper we are concerned with the iR
divergences in  Green functions 
for  non-exceptional Euclidean momenta and not the
additional iR divergences for real processes.)
The problem is to sum them up and extract the
iR behaviour of the Green functions.
We could handle this problem in the following way.
We have shown in Ref.\ \cite{mrs} that for a particular value 
(chosen to each order in $1/N$) of the 
gauge parameter in a specific non-local gauge, the logarithmic iR 
divergences are absent. As a consequence the iR behaviour 
of the Green functions to all orders is known. The 
limiting behaviour is a conformal field theory where
the photon has non-canonical scaling
dimension one for the entire range of $d$, in contrast to the
engineering dimension $(d-2)/2$. The fermion continues to
have the canonical dimension $(d-1)/2$.

This behaviour for the fermions is of course gauge-dependent
and special to this gauge. The Green functions involving only
the photons is gauge invariant and the scaling dimension one
would be valid in any gauge. Our specific choice of gauge
has the advantage of extracting this information without
being cluttered by the powers of logarithms in the intermediate
stages of the calculations.

Thus we know the iR behaviour of the Green functions to all 
orders for a particular choice of the gauge parameter in a 
specific non-local gauge.
It is instructive to know how the logs add up for other values
of gauge parameter and also in other gauges. In this paper, we 
predict the behaviour to all orders without 
detailed calculations. 

This is done using the Zumino identities \cite{zumino} which exactly relate
the Green functions in one choice of gauge to those in another.
Such a relation is also
called the LKF transformation \cite{lkf}; this name has been mostly used for
relation between various conventional covariant gauges (for example, 
between the Landau gauge and the Feynman gauge).
 We are interested in a 
more general class of gauges, including non-local choices. Although the relevant
relation is contained in Zumino's paper \cite{zumino}, we 
rederive it in a way which
is suitable for our purpose.

This far, our discussion has been restricted to a non-local gauge. It is also of interest
to know how the Green functions behave in a conventional gauge
such as the (local) Lorentz gauge. With our choice of the non-local gauge
the longitudinal part of the propagator has the iR behaviour
$q_\mu q_\nu/q^3$, which is softer than that in the Lorentz gauge, 
$q_\mu q_\nu/q^4$. This feature led to only log iR divergences
in fermionic Green functions which further could be cancelled 
to each order by adjusting the gauge parameter at that order.
In contrast, in the Lorentz gauge, the iR divergences become
increasingly worse with the number of loops. A resummation
seems to be beyond reach. We find out the
iR behaviour in this case also using the Zumino identities \cite{other}.

Consider the generating functional of the Green functions in a general
non-local gauge parametrized by function $g(x,y)$\begin{equation}
Z=\frac{\int\exp\ob{-\frac{1}{2}\partial A\cdot g\cdot\partial A+j\cdot A+\bar{\eta}\cdot\psi+\eta\cdot\bar{\psi}}}{\int\exp\ob{-\frac{1}{2}\partial A\cdot g\cdot\partial A}}\,.\label{eq:1}\end{equation}
 Here $\int$ stands for the measure\begin{equation}
\int{\cal D}\psi{\cal D}\bar{\psi}{\cal D}A\;\exp\ob{-S},\label{eq:2}\end{equation}
where $S$ is the gauge-invariant Euclidean action. Also, $j_\mu$ is the source
for the vector potential $A_\mu$, and the Grassmann variables $\eta$ and $\bar\eta$
are the sources for the fermions. We have used the notation
\begin{equation}
\partial A=\partial^{\mu}A^{\mu},\label{eq:3}\end{equation}
while the $\cdot$ operation signifies the inner product in the Hilbert space,
 involving integration over spacetime
variables and summation of discrete (spin and/or internal) labels.
For example, \begin{eqnarray}
j\cdot A=\int d^{3}x\, j^{\mu}(x)A^{\mu}(x)\,,\label{eq:4}\\
\partial A \cdot g \cdot A=\int d^3x d^3y \partial^\mu A^\mu (x) g(x,y)
\partial^\nu A^\nu(y)\,.\end{eqnarray}
 Varying the gauge function $g$ by an infinitesimal amount $\delta g$,
we get \begin{equation}
\delta Z=-\frac{1}{2}\ob{\tb{\partial A\cdot\delta g\cdot\partial A}_{j,\eta,\bar{\eta}}-Z\tb{\partial A\cdot\delta g\cdot\partial A}},\label{eq:5}\end{equation}
where \begin{equation}
\tb{{\cal O}}_{j,\eta,\bar{\eta}}=\frac{\int\exp\ob{-\frac{1}{2}\partial A\cdot g\cdot\partial A+j\cdot A+\bar{\eta}\cdot\psi+\eta\cdot\bar{\psi}}{\cal O}}{\int\exp\ob{-\frac{1}{2}\partial A\cdot g\cdot\partial A}}\label{eq:6}\end{equation}
is the expectation value of operator ${\cal O}$ in presence of sources
$j$, $\eta$ and $\bar{\eta}$. We take $\tb{{\cal O}}$ without any suffix to
denote the corresponding expectation value when the sources are
set to zero. As a consequence of Ward identities,
the correlations involving longitudinal photons in eq.\ (\ref{eq:5}) can 
be related
to pure fermion correlations. This leads to the Zumino identities 
which we now derive in a form suitable for our purpose.

Invariance of $Z$ under the
following change of integration variables (corresponding to an infinitesimal
gauge transformation) in the numerator of eq.\ (\ref{eq:1}) \begin{equation}
\left.\begin{array}{c}
\delta A^{\mu}(x)=-\partial^{\mu}\epsilon(x),\\
\delta\psi(x)=ie\epsilon(x)\psi(x)\quad,\quad\delta\bar{\psi}(x)=-ie\epsilon(x)\bar{\psi}(x)\end{array}\right.
\label{eq:7}\end{equation}
gives the basic Ward identity
\begin{equation}
\tb{\ob{\partial^{2}g\cdot\partial A+\partial j+ie\ob{\bar{\eta}\psi-\eta\bar{\psi}}}_{x}}_{j,\eta,\bar{\eta}}=0\label{eq:8}\end{equation}
 where we have used the notation ${\cal O}_{x}$ to mean ${\cal O}(x)$.
Also,  $\partial^2 g$ can be regarded as the product of the symmetric matrices
$(\partial^2)_{xy}=\delta^{(3)}(x-y)\partial^2_x$ and $g_{xy}$ (note that eq.\ (\ref{eq:1})
picks out the symmetric part of $g$).
Applying the operation $\partial^{\mu}_y({\delta}/\delta j^{\mu}(y))$
on eq.\ (\ref{eq:8}), we get \begin{equation}
\tb{\partial A_{y}\ob{\partial^{2}g\cdot\partial A+\partial j+ie\ob{\bar{\eta}\psi-\eta\bar{\psi}}}_{x}}_{j,\eta,\bar{\eta}}=\ob{\partial_x^{2}\delta_{xy}}Z\label{eq:9}\end{equation}
 where $\delta_{xy}$ stands for the Dirac-delta function $\delta^{(3)}(x-y)$.
Setting the sources to zero in the above equation, we get \begin{equation}
\tb{\partial A_{y}\ob{\partial^{2}g\cdot\partial A}_{x}}=\partial_x^{2}\delta_{xy}.\label{eq:10}\end{equation}
 This is the conventional Ward identity for the longitudinal part
of the photon propagator. From the difference of eq.\ (\ref{eq:9})
and ($Z$ times) eq.\ (\ref{eq:10}), we get \begin{eqnarray}
\tb{\partial A_{y}\ob{\partial A+\ob{\partial^{2}g}^{-1}\cdot\ob{\partial j+ie\ob{\bar{\eta}\psi-\eta\bar{\psi}}}}_{x}}_{j,\eta,\bar{\eta}}
 - \,\, Z\tb{\partial A_{y}\partial A_{x}}=0.\label{eq:11}\end{eqnarray}
 In the last step, we multiplied with the appropriate element of the matrix 
$(\partial^2 g)^{-1}$, which is to be regarded as the inverse of the matrix $\partial^2 g$.
 Using eq.\ (\ref{eq:11}) in eq.\ (\ref{eq:5}), we get \begin{equation}
\delta Z=\frac{1}{2}\tb{\partial A\cdot\delta g\cdot\ob{\partial^{2}g}^{-1}\cdot\ob{\partial j+ie\ob{\bar{\eta}\psi-\eta\bar{\psi}}}}_{j,\eta,\bar{\eta}}.\label{eq:12}\end{equation}

 This is not yet the convenient form for our use. We proceed to eliminate $\partial A$ from eq.\ (\ref{eq:12})
and arrive at an equation involving fermion correlations only. We apply the operation
$\bar{\eta}_{\alpha}(y)(\delta / \delta\bar{\eta}_{\alpha}(y))$
on eq.\ (\ref{eq:8}) to get \begin{eqnarray}
 \tb{\ob{\partial^{2}g\cdot\partial A+\partial j+ie\ob{\bar{\eta}\psi-\eta\bar\psi}}_{x}\ob{\bar{\eta}\psi}_{y}}_{j,\eta,\bar{\eta}}
=\,\, -ie\tb{\ob{\bar{\eta}\psi}_{x}}_{j,\eta,\bar{\eta}}\delta_{xy}\label{eq:13}\end{eqnarray}
 and the operation $\eta_{\alpha}(y)(\delta /\delta\eta_{\alpha}(y))$
to get \begin{eqnarray}
 \tb{\ob{\partial^{2}g\cdot\partial A+\partial j+ie\ob{\bar{\eta}\psi-\eta\bar\psi}}_{x}\ob{\eta\bar{\psi}}_{y}}_{j,\eta,\bar{\eta}}
 =\,\, ie\tb{\ob{\eta\bar{\psi}}_{x}}_{j,\eta,\bar{\eta}}\delta_{xy}.\label{eq:14}\end{eqnarray}
 The difference of eq.\ (\ref{eq:13}) and eq.\ (\ref{eq:14}) yields
\begin{eqnarray}
 \tb{\ob{\partial^{2}g\cdot\partial A+\partial j+ie\ob{\bar{\eta}\psi-\eta\bar{\psi}}}_{x}\ob{\bar{\eta}\psi-\eta\bar{\psi}}_{y}}_{j,\eta,\bar{\eta}}
 =\,\, -ie\tb{\ob{\bar{\eta}\psi+\eta\bar{\psi}}_{x}}_{j,\eta,\bar{\eta}}\delta_{xy}.\label{eq:15}\end{eqnarray}
Multiplying with $(\partial^2 g)^{-1}_{zx}$ and integrating over $x$, we obtain,
\begin{eqnarray}
-\tb{\partial A_{z}\ob{\bar{\eta}\psi-\eta\bar{\psi}}_{y}}_{j,\eta,\bar{\eta}}
&=&\tb{\ob{\ob{\partial^{2}g}^{-1}\cdot\ob{\partial j+ie\ob{\bar{\eta}\psi-\eta\bar{\psi}}}}_{z}\ob{\bar{\eta}\psi-\eta\bar{\psi}}_{y}}_{j,\eta,\bar{\eta}}\nonumber\\
 &&+ie\ob{\partial^{2}g}_{zy}^{-1}\tb{\ob{\bar{\eta}\psi+\eta\bar{\psi}}_{y}}_{j,\eta,\bar{\eta}}.\label{eq:16}\end{eqnarray}
 Also from eq.\ (\ref{eq:8}), \begin{equation}
-\tb{\partial A_{x}}_{j,\eta,\bar\eta}=\tb{\ob{\ob{\partial^{2}g}^{-1}\cdot\ob{\partial j+ie\ob{\bar{\eta}\psi-\eta\bar{\psi}}}}_{x}}_{j,\eta,\bar{\eta}}\label{eq:17}.\end{equation}
 Using eq.\ (\ref{eq:17}) and eq.\ (\ref{eq:16}) in eq.\ (\ref{eq:12}), and using the fact that $(\partial^2 g)^{-1}$ is a
symmetric matrix, we have finally \begin{eqnarray}
 \delta Z&=&\frac{1}{2}\Big[\tb{\ob{\partial j+ie\ob{\bar{\eta}\psi-\eta\bar\psi}}\cdot\delta\ob{\partial^{2}g\partial^{2}}^{-1}\cdot\ob{\partial j+ie\ob{\bar{\eta}\psi-\eta\bar\psi}}}_{j,\eta,\bar{\eta}}\nonumber \\
 &  & -e^{2}\delta\ob{\partial^{2}g\partial^{2}}^{-1}_{00}\tb{\bar{\eta}\cdot\psi+\eta\cdot\bar{\psi}}_{j,\eta,\bar\eta}\Big].\label{eq:18}\end{eqnarray}
 Here we have used 
$\ob{\partial^{2}g}^{-1}\cdot\delta g\cdot\ob{\partial^{2}g}^{-1}
=-\delta\ob{\partial^{2}g\partial^{2}}^{-1}$,
which can be easily checked by going over to the momentum space:
$(-k^2g(k))^{-1} \delta g(k) (-k^2g(k))^{-1} 
=-(1/k^2)\delta (1/g(k)) (1/k^2)$. 
(We will come across explicit examples in momentum space later in this paper.)
Also, $\ob{\partial^{2}g\partial^{2}}_{00}^{-1}$ appears in eq.\ (\ref{eq:18}) as follows:
\begin{eqnarray}
 &  & \int d^{3}x\ob{\ob{\partial^{2}g}^{-1}\cdot\delta g\cdot\ob{\partial^{2}g}^{-1}}_{xx}\tb{\ob{\bar{\eta}\psi+\eta\bar{\psi}}_{x}}\nonumber \\
 && = - \int d^{3}x\delta\ob{\partial^{2}g\partial^{2}}_{xx}^{-1}\tb{\ob{\bar{\eta}\psi+\eta\bar{\psi}}_{x}}\nonumber\\
 && =  -\delta\ob{\partial^{2}g\partial^{2}}_{00}^{-1}\tb{\ob{\bar{\eta}\cdot\psi+\eta\cdot\bar{\psi}}}\label{eq:19} \end{eqnarray}
 since $\ob{\partial^2 g\partial^2}_{xy}^{-1}$ depends only on the difference
$x-y$ for a translation-invariant gauge function $g$.
 
 {\it The dependence of all Green functions
on the function $g_{xy}$ is contained in eq.\ (\ref{eq:18}).} The simplest case of the photon propagator
is obtained by applying the operation $\delta^{2}/(\delta j^{\mu}(x)\delta j^{\nu}(y))$
on eq.\ (\ref{eq:18}) and then setting the sources to zero:\begin{equation}
\delta\Delta_{\mu\nu}(x,y)=\partial_{x}^{\mu}\partial_{y}^{\nu}\delta 
(\partial^2 g \partial^2)^{-1}_{xy}.\label{eq:21}\end{equation}
 This is consistent with the Ward identity eq.\ (\ref{eq:10}).

We now obtain the {\it dependence of the fermion propagator} on the choice of the
gauge function $g$.
Applying $\delta^{2}/(\delta\bar{\eta}_{\gamma}(x)\delta\eta_{\delta}(y))$
to eq.\ (\ref{eq:18}) and setting all the sources to zero, we get \begin{eqnarray}
\delta S_{\gamma\delta}(x,y)
 = -\delta F_{xy}S_{\gamma\delta}(x,y) \end{eqnarray}
 where $F_{xy}$ stands for \begin{equation}
F_{xy}=e^{2}\ob{(\partial^2 g \partial^2)^{-1}_{00}-
                (\partial^2 g \partial^2)^{-1}_{xy}}.
		\label{eq:23}\end{equation}
 Integrating this equation, we relate the fermion propagator evaluated
with two different gauge functions: \begin{eqnarray}
S_{\gamma\delta}(x,y) & = & \exp\left[-\ob{F-F^0}_{xy}\right]S_{\gamma\delta}^{0}(x,y).\label{eq:24}\end{eqnarray}
 Here $S$ and $S^{0}$ stand for the fermion propagators in the gauges
$g$ and $g_{0}$ respectively; $F^{0}$ is related to $g^{0}$ by
eq.\ (\ref{eq:23}).

We have shown in Ref.\ \cite{mrs} that if we choose a particular non-local
gauge \begin{equation}
g=\frac{1}{\alpha}\ob{1+\frac{\mu}{\sqrt{-\partial^{2}}}}\label{eq:25}\end{equation}
(with $\mu=Ne^2/8$) then it is possible to choose the gauge parameter $\alpha$ to each
order in $1/N$ such that there are no logarithmic corrections to
the fermion propagator and other Green functions. As a consequence,
for this particular value of gauge parameter (call it $\alpha_{0}$)
the iR behaviour of the fermion propagator is that of the free theory
with no anomalous dimension:
\begin{equation}
S^{\alpha_{0}}(x,y)\sim\frac{\not\negmedspace x-\not\negmedspace y}{|x-y|^{3}}\,.\label{eq:26}\end{equation}
 The iR behaviour for other values of gauge parameter $\alpha$ within
the same non-local gauge then follows from
eq.\ (\ref{eq:24}) and eq.\ (\ref{eq:25}).
We now have $F_{xy}=\alpha f(x-y)$ where $f$ is formally the matrix
$f_{xy}=e^2((\partial^2(1+\mu/\sqrt{-\partial^2})\partial^2)^{-1}_{00}
        -(\partial^2(1+\mu/\sqrt{-\partial^2})\partial^2)^{-1}_{xy})$.
It is convenient to represent $f$ by the Fourier transform 
\begin{eqnarray}
f(x-y) & = & e^{2}\int\frac{d^{3}k}{(2\pi)^{3}}\frac{\ob{1-e^{ik.(x-y)}}}{k^{2}\ob{k^{2}+\mu k}}\,.\label{eq:28}\end{eqnarray}
[This is obtained by inserting the completeness relation for the momentum eigenstates in
        $<x|(\partial^2(1+\mu/\sqrt{-\partial^2})\partial^2)^{-1}|y>$.
Note that the factor of $(2\pi)^3$ in eq.\ (\ref{eq:28})
is consistent with that in the Fourier transform of $1_{xy}=\delta(x-y)$.]
We now get the propagator for the gauge parameter $\alpha$:
\begin{eqnarray}
S^{\alpha}(x,y) 
  =  \ob{\frac{1}{\lambda(x,y)}}^{\alpha-\alpha_{0}}S^{\alpha_{0}}(x,y)\label{eq:27}\end{eqnarray}
where \begin{equation}
\lambda(x,y)=\exp [f(x-y)].\label{eq:29}\end{equation}
 The integral in eq.\ (\ref{eq:28}) is 
finite at both the ends $k\rightarrow\infty$ and $k\rightarrow 0$.
For $k\rightarrow 0$, finiteness follows from $\mu k\gg k^2$ and 
$\exp(ik\cdot(x-y))\approx 1+ ik\cdot(x-y)$
(actually, by symmetry, it is the $O(k^2)$ term in the exponential which contributes). 
Now, as $|x-y|\rightarrow\infty$,
$k\cdot (x-y)$ is no longer small, and the integral develops a logarithmic divergence as $k\rightarrow 0$.
Thus, $1/|x-y|$ serves as an infrared cutoff for $k$, and for $|x-y|\rightarrow \infty$,
we expect the integral to behave as $\kappa\ln|x-y|$ where $\kappa$ is a constant. 
This can be explicitly verified, and the constant of proportionality extracted, as follows. We have
\begin{eqnarray}
f(x) & = & \frac{e^{2}}{2\pi^{2}}\int_{0}^{\infty}\frac{dk}{k^{2}+\mu k}\ob{1-\frac{\sin (k|x|)}{k|x|}}\nonumber \\
 & = & \frac{e^{2}}{2\pi^{2}\mu}\int_{0}^{\infty}\frac{du}{\rho u^{2}+u}\ob{1-\frac{\sin u}{u}}\label{eq:30}\end{eqnarray}
 where $\rho=1/(\mu|x|)$. Then,
\begin{eqnarray}
\kappa & = & -\lim_{\rho\rightarrow0}\rho\frac{df}{d\rho}\nonumber \\
 & = & \lim_{\rho\rightarrow0}\frac{e^{2}}{2\pi^{2}\mu}\int_{0}^{\infty}\frac{dv}{(v+1)^{2}}\ob{1-\rho\frac{\sin(v/\rho)}{v}}
\nonumber\\
 & = & \frac{e^{2}}{2\pi^{2}\mu}\int_{0}^{\infty}\frac{dv}{(v+1)^{2}}=\frac{4}{\pi^{2}N}\label{eq:31} \end{eqnarray}
which is finite and non-zero.
This gives the iR ($|x|\rightarrow\infty$) behaviour \begin{eqnarray}
f(x) & \sim &\frac{4}{\pi^{2}N}\ln(\mu|x|)\,,\label{eq:32}\\
\lambda(x,y) & \sim & |x-y|^{\frac{4}{\pi^{2}N}}\,.\label{eq:33}\end{eqnarray}
 Thus the iR behaviour of the fermion propagator for an arbitrary
value of gauge parameter $\alpha$ in our non-local gauge is given
by \begin{eqnarray}
S^{\alpha}(x,y) 
\sim \frac{\not\negmedspace x-\not\negmedspace y}{|x-y|^{3+\frac{4\ob{\alpha-\alpha_{0}}}{\pi^{2}N}}}\,.\label{eq:34}\end{eqnarray}
A special case of eq.\ (\ref{eq:34}) is that the power of $|x-y|$ is $3-8/(3\pi^2 N)$ to the leading order in $1/N$
in the Landau gauge, which is obtained by using $\alpha=0$ and $\alpha_0=2/3$ \cite{alpha0}. This value was
obtained earlier by Aitchison et al \cite{other}. However these authors use a different non-local gauge function (the
small momentum limit of our gauge function)
in the LKF transformation equation, and so need to regularize an ultraviolet infinity and also put
an ultraviolet cutoff scale. Our method is free from these complications. 

Eq.\ (\ref{eq:34}) suggests that the iR behaviour for other values of
$\alpha$ is again a CFT, albeit with a non-zero anomalous dimension
for the fermion. We may check this by obtaining the dependence of
the {\it four-fermion Green function} \begin{equation}
S_{\gamma_{1},\gamma_{2};\delta_{1}\delta_{2}}(x_{1},x_{2};y_{1},y_{2})=\left.\frac{\delta^{4}Z}{\delta\bar{\eta}_{\gamma_{1}}(x_{1})\delta\bar{\eta}_{\gamma_{2}}(x_{2})\delta\eta_{\delta_{1}}(y_{1})\delta\eta_{\delta_{2}}(y_{2})}\right|_{j=\eta=\bar{\eta}=0}\label{eq:35}\end{equation}
on $\alpha$.
From eq.\ (\ref{eq:18}) we get \begin{eqnarray}
\delta S_{\gamma_{1},\gamma_{2};\delta_{1}\delta_{2}}(x_{1},x_{2};y_{1},y_{2}) & = & \left[\delta\left(F_{x_{1}x_{2}}+F_{y_{1}y_{2}}-F_{x_{1}y_{2}}-F_{x_{2}y_{1}}\right.\right.\nonumber\\
 &  & \left.\left.-F_{x_{1}y_{1}}-F_{x_{2}y_{2}}\right)\right]S_{\gamma_{1},\gamma_{2};\delta_{1}\delta_{2}}(x_{1},x_{2};y_{1},y_{2}).\end{eqnarray}
 The solution to this equation can be cast in the form \begin{eqnarray}
 &  & S_{\gamma_{1},\gamma_{2};\delta_{1}\delta_{2}}^{\alpha}(x_{1},x_{2};y_{1},y_{2})\nonumber\\
 & = & \rb{\frac{\lambda(x_{1},x_{2})\lambda(y_{1},y_{2})}{\lambda(x_{1},y_{1})\lambda(x_{1},y_{2})\lambda(x_{2},y_{1})\lambda(x_{2},y_{2})}}^{\alpha-\alpha_{0}}S_{\gamma_{1},\gamma_{2};\delta_{1}\delta_{2}}^{\alpha_{0}}(x_{1},x_{2};y_{1},y_{2}).
\label{eq:36} \end{eqnarray}
Using eq.\ (\ref{eq:33}), we may write \begin{eqnarray}
 &  & S_{\gamma_{1},\gamma_{2};\delta_{1}\delta_{2}}^{\alpha}(x_{1},x_{2};y_{1},y_{2})\nonumber\\
 & = & \frac{1}{|x_{1}-y_{1}|^{\frac{4(\alpha-\alpha_{0})}{\pi^{2}N}}|x_{2}-y_{2}|^{\frac{4(\alpha-\alpha_{0})}{\pi^{2}N}}}\ob{\frac{\rho}{\eta}}^{\frac{4(\alpha-\alpha_{0})}{\pi^{2}N}}S_{\gamma_{1},\gamma_{2};\delta_{1}\delta_{2}}^{\alpha_{0}}(x_{1},x_{2};y_{1},y_{2})\label{eq:37} \end{eqnarray}
 for the iR behaviour. Here $\rho$ and $\eta$ are the conformal
invariant cross-ratios \begin{eqnarray}
\rho=\frac{|x_{1}-x_{2}||y_{1}-y_{2}|}{|x_{1}-y_{1}||x_{2}-y_{2}|} & , & \eta=\frac{|x_{1}-y_{2}||x_{2}-y_1|}{|x_{1}-y_{1}||x_{2}-y_{2}|}\,.\label{eq:38}\end{eqnarray}
As $S^{\alpha_{0}}$ has a structure required
by conformal invariance (in the infrared), eq.\ (\ref{eq:37}) implies that
$S^{\alpha}$ also has
such a structure \cite{gatto}. 
This is consistent with an anomalous
dimension $4(\alpha-\alpha_{0})/(\pi^{2}N)$ for the fermion. It is interesting
that the angular dependence of the scattering amplitude is modified
by a simple factor given by a power of $\rho/\eta$ when one changes
the gauge parameter.

We now address the gauge dependence of the
{\it three-point fermion-photon Green function}
\begin{equation}
V_{\mu;\gamma,\delta}(z;x,y)=\left.\frac{\delta^{3}Z}{\delta j^{\mu}(z)\delta\bar{\eta}_{\gamma}(x)\delta\eta_{\delta}(y)}\right|_{j,\eta,\bar{\eta}=0}.\label{eq:V1}\end{equation}
 From eq.\ (\ref{eq:18}) and eq.\ (\ref{eq:V1}) it follows that \begin{eqnarray}
\delta V_{\mu;\gamma,\delta}(z;x,y) & = & -\delta F(x,y) V_{\mu;\gamma,\delta}(z;x,y)\nonumber\\
 &  & +(i/e)\partial_{z}^{\mu}\ob{\delta F(z,x)-\delta F(z,y)}
 S_{\gamma\delta}(x,y).\label{eq:V2} \end{eqnarray}
 For the part of the 3-point function corresponding to a longitudinal
photon, we get a simpler equation by applying $\partial^\mu_z$ on eq.\ (\ref{eq:V2}): \begin{eqnarray}
\delta\partial_{z}^{\mu} V_{\mu;\gamma,\delta}(z;x,y) & = & -\delta F(x,y)\partial_{z}^{\mu} V_{\mu;\gamma,\delta}(z;x,y)\label{eq:V2a}\\
 &  & -ie\ob{\delta\ob{\partial^{2}g}_{zx}^{-1}-\delta\ob{\partial^{2}g}_{zy}^{-1}}S_{\gamma\delta}(x,y).\nonumber \end{eqnarray}
 This is satisfied by the Ward identity \begin{equation}
\partial_{z}^{\mu} V_{\mu;\gamma,\delta}(z;x,y)=-ie\rb{\ob{\partial^{2}g}_{zx}^{-1}-\ob{\partial^{2}g}_{zy}^{-1}}S_{\gamma\delta}(x,y)\label{eq:V3}\end{equation}
 following from eq.\ (\ref{eq:8}). 
 On the other hand for the part $\widetilde V$ of $V$ that relates to a transverse
photon, \begin{equation}
\widetilde{V}_{\mu;\gamma,\delta}(z;x,y)=
\ob{\eta_{\mu\nu}-\frac{\partial_\mu^z \partial_\nu^z}{\partial^2_z}}{V}_{\nu;\gamma,\delta}(z;x,y)
,\label{eq:V4}\end{equation}
 eq.\ (\ref{eq:V2}) leads to  \begin{equation}
\delta\widetilde{V}_{\mu;\gamma,\delta}(z;x,y)=-\delta F(x,y)\widetilde{V}_{\mu;\gamma,\delta}(z;x,y).\label{eq:V5}\end{equation}
 Its solution is \begin{equation}
\widetilde{V}^\alpha_{\mu;\gamma,\delta}(z;x,y)=\ob{\frac{1}{\lambda(x,y)}}^{\alpha-\alpha_{0}}\widetilde{V}^{\alpha_0}_{\mu;\gamma,\delta}(z;x,y),\label{eq:V6}\end{equation}
 which is consistent with an anomalous dimension as given in eqs.\ (\ref{eq:27}) and (\ref{eq:34})
 for the fermion and
a gauge-invariant anomalous dimension for the photon.

We have shown that the iR behaviour in a class of non-local gauges
parametrized by a parameter $\alpha$ is given by a CFT with the fermion
anomalous dimension depending on the parameter $\alpha$. Using this
we now obtain {\it the iR behaviour in the usual class of local gauges}.
This turns out to be very instructive regarding attempts to resum
iR divergences of the perturbation theory. 

We obtain the fermion propagator in the local gauge corresponding
to the Lorentz gauge term $-(1/(2\alpha))\ob{\partial A}^{2}$
by comparing with that for the non-local gauge
\bea
-\frac{1}{2\alpha}\ob{\partial A}\cdot\ob{1+\frac{\mu}{\sqrt{-\partial^{2}}}}\cdot\ob{\partial A}
\eea
 with the same gauge parameter $\alpha$. For the local Lorentz
gauge \begin{eqnarray}
F^{L}(x,y)
 & = & \alpha e^2\int\frac{d^{3}k}{(2\pi)^{3}}\frac{1}{k^{4}}\ob{1-e^{ik.(x-y)}}\nonumber \\
& = & \frac{\alpha e^{2}}{2\pi^{2}}\int_{0}^{\infty}\frac{dk}{k^{2}}\ob{1-\frac{\sin (k|x-y|)}{k|x-y|}}\nonumber \\
 &=& \frac{\alpha e^2}{2\pi^2}|x-y| 
\int_{0}^{\infty}du\frac{1}{u^{2}}
 \ob{1-\frac{\sin u}{u}}\,.
\label{eq:V7}\end{eqnarray}
The integral can be evaluated by rewriting the integrand as $(u-\sin u)(1/u^3)$, integrating by parts twice, and
using $\int_0^\infty du(\sin u/u)=\pi/2$. We thus find $F^L(x,y)=\alpha e^2 |x-y|/(8\pi)$.
Using eq.\ (\ref{eq:24}) the iR behaviour in the local 
gauge then comes out as \begin{equation}
S^{L}(x,y)\sim\exp\rb{-\frac{\alpha e^2}{8\pi} |x-y|}\frac{\not\negmedspace x-\not\negmedspace y}{|x-y|^{3-\frac{4\alpha_{0}}
{\pi^{2}N}}}\,.\label{eq:V8}\end{equation}
[It may be noted that starting from our non-local gauge, we reach the local Lorentz gauge through infinitesimal
changes of the gauge function by (formally) varying  $\mu$ from $Ne^2/8$ to zero, at a fixed value of $\alpha$.
As $\mu$ is decreased, we pass from the $\mu k\gg k^2$ regime to the $k^2\gg\mu k$ regime, and finally
reach $F^L(x,y)$ smoothly.] 

Now $\alpha>0$ for the contribution of the gauge-fixing term to be of the correct sign to make the Euclidean
functional integral converge.
Thus eq.\ (\ref{eq:V8}) tells us that the fermion propagator falls off exponentially as if the fermion 
has developed a mass $\alpha e^2/(8\pi) $! 
However, this apparent mass is spurious, since fermion mass cannot be dynamically generated in
perturbation theory (indeed, 
the propagator of eq.\ (\ref{eq:V8}) is 
proportional to $\rlap/p$ in momentum space).

This strikingly illustrates the pitfalls in resumming iR divergences in
perturbation theory. There are severe iR divergences in the local Lorentz gauge,
because the longitudinal part of the photon propagator has a $1/k^2$ behaviour
in the iR. The cumulative effect is an apparent mass term in
gauge non-invariant Green functions. The apparent mass is a gauge artifact; it does not
appear in gauge-invariant correlation functions.

In this paper, we determined how the iR logarithms of massless QED$_3$ add up
for arbitrary values of the gauge parameter in a non-local gauge and also in the usual
Lorentz gauge. 
We demonstrated by studying various correlation functions that the iR behaviour in the non-local gauge
is that of a CFT with a continuously varying anomalous dimension for the fermion. We also demonstrated the
pitfalls in summing the severe iR divergences of the usual Lorentz gauge (the fermion propagator falls
off exponentially as if there is a fermion mass, which is actually a gauge artifact); thus it is the 
non-local gauge which is suitable for studying the iR behaviour of this theory. 
The implications of the calculation in the non-local gauge for the important issue of 
anomalous dimension of
the gauge-invariant dressed fermion will be presented elsewhere \cite{mrs2}.

\section*{Acknowledgements}
We thank A. Bashir and A. Davydychev for useful communications. I.M. thanks IMSc, Chennai for hospitality
during the course of this work.


\begin{thebibliography}{10}
\bibitem{alpha0}
$\alpha$ is related to the gauge parameter $\xi$ used in Ref.\ \cite{mrs} by
$\alpha=1-\xi$.
\end{thebibliography}
\end{document}